# The rule for a subdiffusive particle in an extremely diverse environment

Ophir Flomenbom

*Flomenbom-BPS, 13900 Marquesas Way, Marina del Rey, CA 90292*

**Abstract** The dynamics of a subdiffusive continuous time random walker in an inhomogeneous environment is analyzed. In each microscopic jump, a random time is drawn from a waiting time probability density function (WT-PDF) that decays as a power law: $\varphi(t;k) \sim \frac{k}{(1+kt)^{1+\beta}}$, $0 < \beta \leq 1$. The parameter $k$, which is the diffusion coefficient for the jump, is a random quantity also; in each jump, the value of $k$ is drawn from a PDF, $p(k) \sim \frac{1}{\tilde{k}}\left(\frac{\tilde{k}}{k}\right)^\gamma$, $0 \leq \gamma < 1$, and $0 < k \leq \tilde{k}$, for a cutoff parameter $\tilde{k}$. We show that this system exhibits a transition in the scaling law of its effective WT-PDF, $\psi(t)$, which is obtained when averaging $\varphi(t;k)$ with $p(k)$. $\psi(t)$ decays as a power law, $\psi(t) \sim \frac{1}{t^{1+\mu}}$, and $\mu$ is given by two different formula. When $1-\gamma > \beta$, $\mu = \beta$, but when $1-\gamma < \beta$, $\mu = 1-\gamma$. The rule for the scaling of $\psi(t)$ reflects the competition between two different mechanisms for subdiffusion: subdiffusion due to the heavily tailed $\varphi(t;k)$ for microscopic jumps, and subdiffusion due to the collective effect of an environment made of many slow local regions. These two different mechanisms for subdiffusion are not additive, and compete each other. The reported transition is dimension independent, and disappears when the power $\beta$ is also distributed, in the range, $0 < \beta \leq 1$. Simulations exemplified the transition, and implications are discussed.

*Introduction.-* The statistical properties of random walks have long been used for explaining processes in physics, chemistry and biology [1-11]. Among the basic properties of random walks is the scaling of the mean square displacement with time. When each microscopic jump of a symmetric random walker is associated with an exponential waiting time probability density function (WT-PDF) with the same rate $k$, the mean square displacement of the random walker is proportional to the time the process has been going on. This is the well-known diffusive behavior, and the proportionality constant is the diffusion constant. Diffusion plays an important role in characterizing complex process, e.g. reactions [2-9]. In many systems in condensed phase, the environment is heterogeneous. This can be modeled by taking the value of the rate $k$ to depend on the local environment. For example, a model for a random walk in a constantly changing environment is one in which each microscopic transition has a random value for the rate $k$ drawn from a PDF $p(k)$, e.g. [9]. This simple model for a continuous time random walk in a random environment can lead to a different scaling law for the mean square displacement with time. For certain choices of $p(k)$, this model can lead to subdiffusion, i.e. the random walker's mean square displacement scales sub-linearly with time. This subdiffusive behavior can also be detected by constructing the effective WT-PDF for individual transition, denoted by $\psi(t)$. $\psi(t)$ is obtained by averaging the underlying dynamics with the environment. For a subdiffusive process, $\psi(t)$ decays as an inverse power law with an infinite mean, i.e. $\psi(t) \sim t^{-1-\beta}$, $0 < \beta \leq 1$, and the mean square displacement scales as a power-law with an exponent $\beta$.

Here, we analyze a model for a symmetric random walker in a random environment in which the underlying microscopic waiting times are distributed as an inverse power law WT-PDF with an infinite mean, characterized by a power $\beta$ and a rate parameter $k$; see Eq. (5). The environment, i.e. the distribution of the $k$s, is characterized by a power law function with a power $\gamma$, see Eq. (2). We calculate the effective WT-PDF for individual jumps of such a system, and show that this system exhibits a transition in the scaling of the effective WT-PDF for individual jumps. The effective WT-PDF always decays as a power law with an infinite mean, $\psi(t) \sim t^{-1-\mu}$, $0 < \mu \leq 1$, but the formula for $\mu(\gamma, \beta)$ changes in a critical point, $\beta = 1 - \gamma$. Importantly, this is an indication for a change in the mechanism that controls the dynamics. When $\beta$ is larger than $1 - \gamma$, $\mu = 1 - \gamma$, meaning that the collective effect of the environment becomes the mechanism that controls the dynamics. In this regime, the collective effect of an environment made of many slow regions leads to a slower subdiffusion than the subdiffusion due to the microscopic dynamics. When $\beta$ is smaller than $1 - \gamma$, $\mu = \beta$, and the slow individual transitions control the dynamics. Within the model analyzed here, the transition is independent of the dimension of the system, but vanishes when the power $\beta$ is distributed also. Relationships of the reported transition to dynamics on a network made of heterogeneous fractals, dynamics of single file systems and dynamics in crowded biological cells are discussed.

*Normal dynamics in random environments.-* A well known result in stochastic processes gives the scaling of the mean square displacement of a normal diffusive symmetric random walker as a linear function of the time, $<r^2> \sim D_0 t$. In this paper, the sign ~ symbolizes asymptotic scaling. The diffusion coefficient, $D_0$, can be related to the temperature, the viscosity of the medium and the particle's size, by the Einstein-Stokes relation, e.g. [6]. The linear scaling of $<r^2>$ with time is pretty general: it holds in any dimension and may hold also when the environment is inhomogeneous (while adjusting the value of $D_0$ so it reflects average properties of the environment). However, there are cases in which the environment leads to a very slow diffusion, locally, and when there are many such slow local regions, the overall scaling for $<r^2>$ can be slower than a square root. A simple model for subdiffusion originates from a combination of a normal diffusive random walker with an exponential WT-PDF for individual jumps [9],

$$\varphi(t;k) = ke^{-kt}, \tag{1}$$

and a random environment that leads to a distribution in $k$ with many small $k$s; namely, the value for $k$ is a random quantity, and it is drawn in each transition from a PDF,

$$p(k) \sim \frac{1}{\tilde{k}}\left(\frac{\tilde{k}}{k}\right)^{\gamma}, 0 \leq \gamma < 1, \tag{2}$$

and $k$ is between zero and $\tilde{k}$, for some cut-off parameter $\tilde{k}$ (the fastest timescale in the system is bounded from below). For this model, the effective WT-PDF of the random walker, $\psi(t)$, which is defined by,

$$\psi(t) = \int_0^\infty p(k)\varphi(t;k)dk, \tag{3}$$

decays as a power law, e.g. [9],

$$\psi(t) \sim \frac{\tilde{k}}{(\tilde{k}t)^{1+\mu}} \quad ; \quad \mu = 1 - \gamma. \tag{4}$$

A consequence of a heavily tailed $\psi(t)$ is that the scaling for $<r^2>$ is sub-linear in time, e.g. [7],

$$<r^2> \sim (\tilde{D}_0 t)^{\mu}.\text{*}$$

---

* This result for the mean absolute displacement is obtained from the general scaling-law for a CTRW [in this paper $\bar{g}(s) = \int_0^\infty g(t)e^{-st}dt$ is the Laplace transform of the function $g(t)$]:

$$<\overline{r^2}(s)> \sim \frac{\bar{\psi}(s)/s}{1-\bar{\psi}(s)}$$

$\widetilde{D}_0$ is a parameter. This model shows how the environment can affect the diffusion, leading to subdiffusion. Here, the reason for the subdiffusion is that there are many regions in which the local rate $k$ is very small. As $\gamma \to 0$, the diffusion becomes normal.

***Subdiffusive dynamics in random environments.-*** In this paper, we extend the above well-known simple model for normal diffusion in a random environment involves *subdiffusion* in a random environment. We consider a model in which the waiting times for individual jumps are distributed according to a power law PDF, all having the same scaling,

$$\varphi(t;k) \sim \frac{k}{(1+kt)^{1+\beta}}, \quad 0 < \beta \leq 1. \tag{5}$$

In each jump the parameter $k$ is drawn from the PDF in Eq. (2); namely, the random environment is modeled in the same way as before. The calculations compute the effective WT-PDF, given by the integral equation,

$$\psi(t) = \tilde{k}\left(\frac{1}{\tilde{k}t}\right)^{2-\gamma} \int_0^{\tilde{k}t} \frac{s^{1-\gamma}}{(1+s)^{1+\beta}} ds.$$

The integral $\int_0^{\tilde{k}t} \frac{s^{1-\gamma}}{(1+s)^{1+\beta}} ds$ is approximated by the sum of two integrals,

$$\int_0^{\tilde{k}t} \frac{s^{1-\gamma}}{(1+s)^{1+\beta}} ds \lesssim \int_0^1 s^{1-\gamma} ds + \int_1^{\tilde{k}t} s^{-\gamma-\beta} ds,$$

with the solution,

$$\int_0^{\tilde{k}t} \frac{s^{1-\gamma}}{(1+s)^{1+\beta}} ds \lesssim \frac{1}{2-\gamma} + \frac{(\tilde{k}t)^{1-\gamma-\beta}-1}{1-\gamma-\beta}.$$

For large *t*, the outcome of the integral depends on the sign of $1 - \gamma - \beta$:

---

and using Tauberian theorem (e.g. [7]) to obtain the Laplace transform of the power law $\psi(t)$.

$$\int_0^{\tilde{k}t} \frac{s^{1-\gamma}}{(1+s)^{1+\beta}} ds \lesssim \begin{cases} \frac{1}{2-\gamma} + \frac{1}{\gamma+\beta-1} & 1-\gamma < \beta \\ \frac{(\tilde{k}t)^{1-\gamma-\beta}}{1-\gamma-\beta} & 1-\gamma > \beta \end{cases},$$

and consequently,

$$\tilde{k}^{-1}\psi(t) \sim \begin{cases} \left(\frac{1}{\tilde{k}t}\right)^{2-\gamma} & 1-\gamma < \beta \\ \left(\frac{1}{\tilde{k}t}\right)^{1+\beta} & 1-\gamma > \beta \end{cases} \tag{6.1}\tag{6.2}$$

Equations (6) predict a transition in the scaling law of the effective WT-PDF for a subdiffusive random walker in a random environment, occurring when $1 - \gamma = \beta$. It is very easy to exemplify this transition in a simulation. Figures 1-3 show the average WT-PDF $\psi(t)$ for different values of $\gamma$ and $\beta$. We take for $\beta$, $\beta = 0.25, 0.5, 0.75$, and each value of $\beta$ is conjugated with two different $\gamma$s, one that leads to Eq. (6.1) and another that leads to Eq. (6.2). The transition is apparent in all cases. Note that this transition in the scaling of the effective WT-PDF is independent of the dimension of the system. The effective WT-PDF depends only on the input WT-PDF for individual transitions and the heterogeneity in the system, and both, in this model, are independent of the dimension of the system.

What about a random environment that leads to a distribution in the power $\beta$ in Eq. (5)? For this case, *no* transition in the scaling law of $\psi(t)$ is observed, with or without an additional averaging over the parameter $k$, as long as the average over the power $\beta$ is done in the range, $\epsilon \leq \beta \leq 1$ (here, $\epsilon \to 0^+$). In particular, when choosing,

$$g(\beta) \sim \beta^{-J} \quad ; \quad 0 \leq J < 1,$$

straightforward calculations give,

$$\psi(t) \sim \frac{k}{(1+kt)^{1+\epsilon}} \left(\frac{1}{\log(1+kt)}\right)^{1-J},$$

and,

$$\psi(t) \sim \frac{\tilde{k}}{(\tilde{k}t)^{1+\epsilon}} \left(\frac{1}{\log(\tilde{k}t)}\right)^{1-J},$$

before and after an additional averaging with $p(k)$ is done, respectively. When choosing,

$$g(\beta) \sim (1-\beta)^{-J}, 0 \leq J < 1,$$

calculations show that,

$$\psi(t) \sim \frac{k}{(1+kt)^2} (\log(1+kt))^J,$$

and,

$$\psi(t) \sim \frac{\tilde{k}}{(\tilde{k}t)^{2-\gamma}} \left(\log(\tilde{k}t)\right)^J,$$

before and after an additional averaging with $p(k)$ is done, respectively. Note that the limit $J \to 0$ should be taken in the former scaling of $\psi(t)$, because the results for the second case are obtained by exploiting the fact that most of the probability is located near one for $g(\beta) \sim (1-\beta)^{-J}$.

The above calculations were done by averaging over the power $\beta$ and then over the parameter $k$. We can reverse the order of the averaging, and first average over the parameter $k$ and then over the $\beta$s. This order of averaging requires two different integration regimes, as can be seen from Eq. (6). One must distinguish between the following two different random environments: in one set of random systems, $\beta$ is in the range, $1 - \gamma \leq \beta \leq 1$, whereas in the other set of random systems $\beta$ is in the range, $0 \leq \beta \leq 1 - \gamma$. When starting with Eq. (6) and performing the restricted averaging with $g(\beta) \sim (1-\beta)^{-J}$, one gets only *one* result for $\psi(t)$ in both averaging regimes, $\psi(t) \sim \frac{\tilde{k}}{(\tilde{k}t)^{2-\gamma}}$. That is, the transition disappears for this choice of $g(\beta)$. Restricted integrations are important when averaging with $g(\beta) \sim \beta^{-J}$. Here, when $\beta$ is integrated over the range $1 - \gamma$ and 1 one gets,

$$\psi(t) \sim \frac{\tilde{k}}{(\tilde{k}t)^{2-\gamma}},$$

but when $\beta$ is integrated over the range 0 and $1 - \gamma$ one gets,

$$\psi(t) \sim \frac{\tilde{k}}{(\tilde{k}t)^{1+\epsilon}} \frac{1}{[\log(\tilde{k}t)]^{1-J}}.$$

Thus, when the environment allows any value of the power $\beta$, the value of $\gamma$ cannot affect the scaling of $\psi(t)$, and no transition can be obtained. However, when the range of allowed $\beta$ values is correlated with the value of $\gamma$, different scaling in $\psi(t)$ can be obtained depending on the shape of $g(\beta)$ and on the integration range for $\beta$. This is an artificial transition because one must prepare the system to fulfill the condition on $\beta$, and this special preparation leads to a different scaling in $\psi(t)$. The physical origin of this artificial transition is the same as in the *fixed $\beta$* case.

***Discussion.-*** How to interpret this transition? Looking on Eq. (6.1), we note that it scales as Eq. (4). Also, Eq. (6.2) scales as Eq. (5). Namely, the effective WT-PDF is determined by the slower mechanism: anomaly due to individual jumps or due to the environment. We also conclude that the underlying subdiffusive dynamics competes with the anomaly due to the heterogeneity in the system. When each transition is very slow relative to the anomaly related to the environment $(1 - \gamma > \beta)$, the diffusion does not depend on the environment heterogeneity. However, when the environment becomes very slow $(1 - \gamma < \beta)$, the individual transitions are, effectively, exponentially distributed from the environment 'perspective', and only the exponent of $p(k)$ enters in the scaling of $\psi(t)$ as if the average over $\varphi(t; k)$ is performed with an exponential WT-PDF for individual transitions.

What are the implications of the reported transition? There are several systems that can be described by this model. For example, a continuous time random walker in a branched lattice. The branches are fractals with the same spectral dimension (the spectral dimension determines the power $\beta$ [10]), but differ in conductivity; namely, each fractal has a transition rate for individual transition drawn from $p(k)$. Here, the statistical properties of the random walker can be tuned by adjusting, for example, the distribution of the fractals' conductivity while keeping fixed their spectral dimension.

Looking beyond a single random walker model, we predict that traces of the transition reported here will be found in a single file system of *N* sub-diffusive hard rods (no bypassing is allowed upon collisions) in a (quasi) one-dimensional geometry [11]. In the sub-diffusive single file, the WT-PDF for individual transitions is given by Eq. (5) for all the particles in it, but the hard rods are not identical and each has a diffusion coefficient *k* drawn from the PDF in Eq. (2). We conjecture that a transition in the scaling of the tagged particle's mean square displacement in this single file will be observed when the collective effect due to the distribution in *k*s leads to a slower dynamics relative to the anomaly attributed to the heavily tailed WT-PDF for individual transitions. As shown here, these two factors are not additive but compete each other.

The last example we consider here is diffusion in living cells [12]. The cell content is dense, but it is also non-uniform (the density fluctuates in space and time). It is well known that crowding is observed in the cell due to its high density content; that is, slow dynamics are observed. The model presented here takes into account both the slow individual transition and the heterogeneity in the environment, and may be found useful in the analysis of the dynamics in living cells. For example, consider a local dynamics governed by an exponential WT-PDF,

$$\varphi(t;k) = ke^{-kt},$$

in which the rate $k$ is related to an energy barrier by the Kramers rate,

$$k = k_0 e^{-\Delta/\tilde{\Delta}},$$

and the energy barrier $\Delta$ is distributed according to,

$$p(\Delta) = \frac{1}{\tilde{E}} e^{-\Delta/\tilde{E}}.$$

This model is equivalent to the model given by Eqs.(1)-(2), and thus the local dynamics obeys,

$$\varphi(t;k) \sim \frac{k_0}{(k_0 t)^{1+\beta}} \quad ; \quad \beta = \frac{\tilde{\Delta}}{\tilde{E}}.$$

The parameter $\tilde{\Delta}$ is related to the temperature, and so it is a constant in the cell environment. The parameter $k_0$ can be related to the local 'friction' and thus can be distributed on the level of the cell. The parameter $\tilde{E}$ determines the scale of the local barriers. It can be argued that although in principle both $\tilde{E}$ and $k_0$ can be distributed, $\tilde{E}$ is more likely to change much less in a changing environment (the nature of the interactions between molecules in the cell is pretty much the same regardless of the local density) and that the local density affects mainly the local friction namely the prefactor, $k_0$. When modeling the distribution in the $k_0$ by a PDF of the form of Eq. (2), we end up with a subdiffusion in a non-uniform environment that can lead to the transition predicted by Eqs. (6). Thus, crowing can affect the dynamics in the cell in two ways: from side it gives rise to the local slow dynamics because dynamics are possible only when many 'microscopic' events occur, and from the other side, it may affect the scaling of the effective WT-PDF, when there are enough 'very dense' local environments. Finally, we note that although the anomalous behavior can be terminated by an exponential 'cut-off', the above analysis refers to a situation in which the anomaly is observed in a time window important to the biological activity.

*Concluding Remarks.-* In this paper, we considered a *subdiffusive* continuous time random walker (characterized by a power $\beta$) in an inhomogeneous environment (characterized by a power $\gamma$). We showed that this system exhibits a transition in the scaling law of its effective WT-PDF, $\psi(t)$. $\psi(t)$ decays as a power law, $\psi(t) \sim \frac{1}{t^{1+\mu}}$, and $\mu$ is given by two different formula. When $1 - \gamma > \beta$, $\mu = \beta$, but when $1 - \gamma < \beta$, $\mu = 1 - \gamma$. The transition in the scaling of $\psi(t)$ reflects the competition between two different mechanisms for subdiffusion: subdiffusion due to the heavily tailed WT-PDF for microscopic jumps, and subdiffusion due to the collective effect of an environment made of many slow local regions. These two different mechanisms for subdiffusion are not additive, and compete each other. The reported transition is dimension independent, and disappears when the power $\beta$ is also distributed in the range, $0 < \beta \leq 1$. We introduced several systems for which this phenomenon can be related, e.g. single file dynamics and diffusion in dense living cells.

**Figure Captions**

Fig 1 A log-log plot of the effective WT-PDF versus time. In each panel, shown are also two curves that correspond to power-laws, $t^{-1-\mu}$, with different values for $\mu$. The left panel shows that the effective WT-PDF scales as the input WT-PDF when $1 - \gamma > \beta$, where the right panel emphasizes the transition, and shows that the scaling of effective WT-PDF is determined by the environment, when $1 - \gamma < \beta$.

Fig 2 The same as Fig 1 with different values for $\gamma$ and $\beta$.

Fig 3 The same as Fig 1 with different values for $\gamma$ and $\beta$.

Figure 1

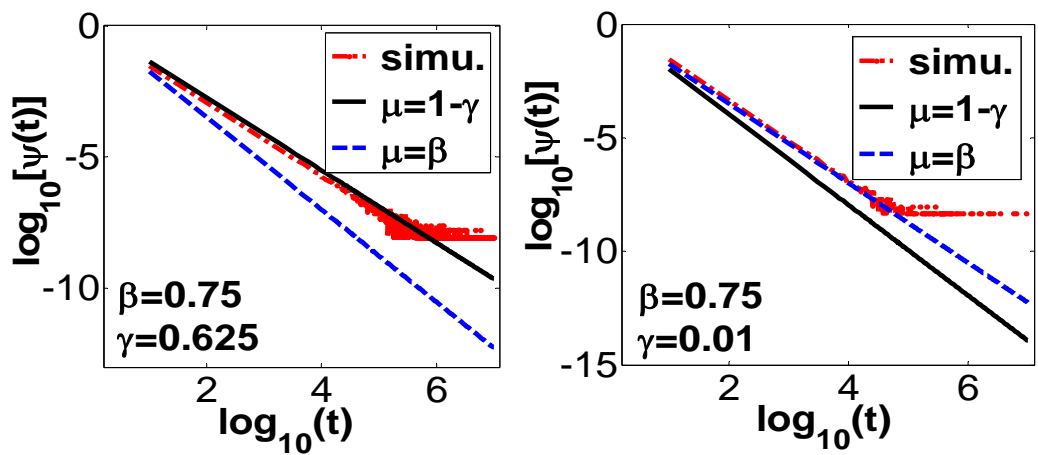

Figure 2

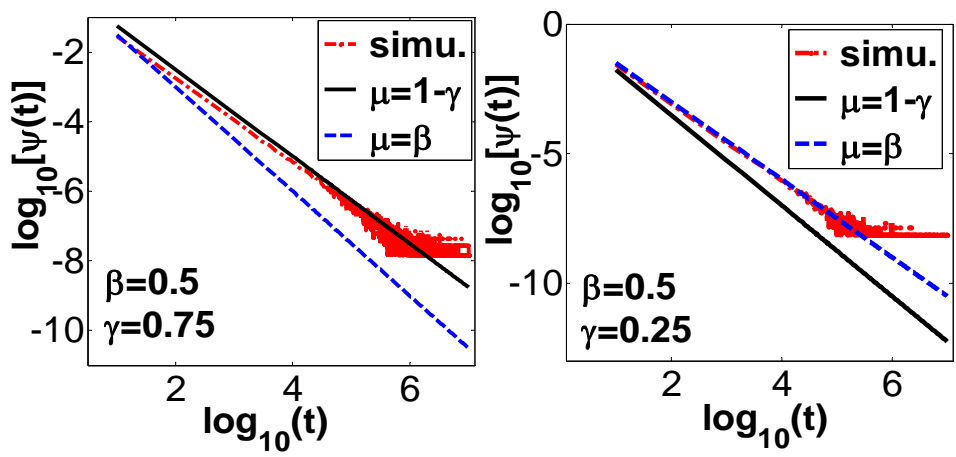

Figure 3

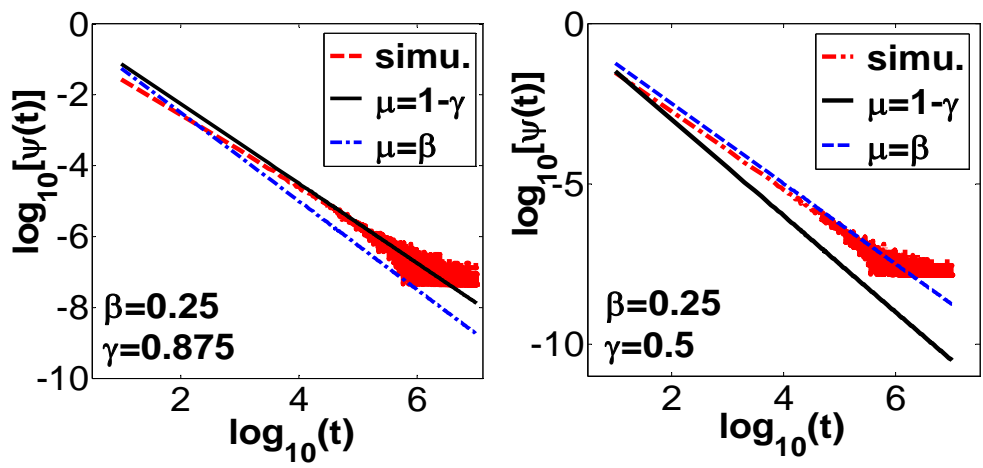

15